\documentclass[
    ,final            
  ]
  {aipproc}

\layoutstyle{8x11single}

\begin{document}

\title{Multipole Extraction: A novel, model independent method}


\author{E. Stiliaris}{
  address={University of Athens
  and Institute of Accelerating Systems \& Applications, Athens, GREECE}
}

\author{C.N. Papanicolas}{
  address={University of Athens
  and Institute of Accelerating Systems \& Applications, Athens, GREECE}
}

\begin{abstract}
A novel method for extracting multipole amplitudes in the nucleon
resonance region from electroproduction data is presented. The
method is based on statistical concepts and it relies heavily on
Monte Carlo and simulation techniques; it produces precise
identification and  determination of the contributing multipole
amplitudes in the resonance region and for the first time a
rigorous determination of the associated experimental
uncertainty. The results are demonstrated to be independent of
any model bias. The method is applied in the reanalysis of the
$Q^{2}=0.127\ GeV^2/c^2$  Bates and Mainz $N\rightarrow \Delta$
data.
\end{abstract}

\maketitle



\section{Motivation}

A significant component of hadronic physics research focuses  on
the understanding of the excitation spectrum of the  proton, the
only stable hadron~\cite{NSTAR01}. Impressive progress has been
achieved in the last decade largely driven by the advances in
accelerator and instrumentation technologies. The high quality of
data that have emerged and substantial progress in theory have
brought a new level of sophistication to the field. Conjectures,
such as the deformation of the nucleon~\cite{gl79} have been
confirmed (see  ~\cite{cnp03} and references therein) and
transformed into an extensive program examining not its existence
but rather the mechanisms that generate it. For the first time
contact with QCD through lattice~\cite{AlexLattice} calculations
and chiral field effective theories~\cite{MVdH,PVY06} has been
established. A most fertile ground of research has been the
identification of the nucleon resonances, the isolation and
interpretation of the contributing multipoles to their excitation
spectrum. Progress has been substantial but slow, due to the
large width of the resonances, their overlapping nature and the
incompleteness of the data base in terms of observables.
Extensive work concerns the investigation of the
$\Delta^{+}(1232)$, the first excited and only isolated state of
the proton,  through the $N\rightarrow \Delta$
reaction~\cite{cnp03,amb03}.

\begin{figure}[h]
  \includegraphics[height=5.8cm]{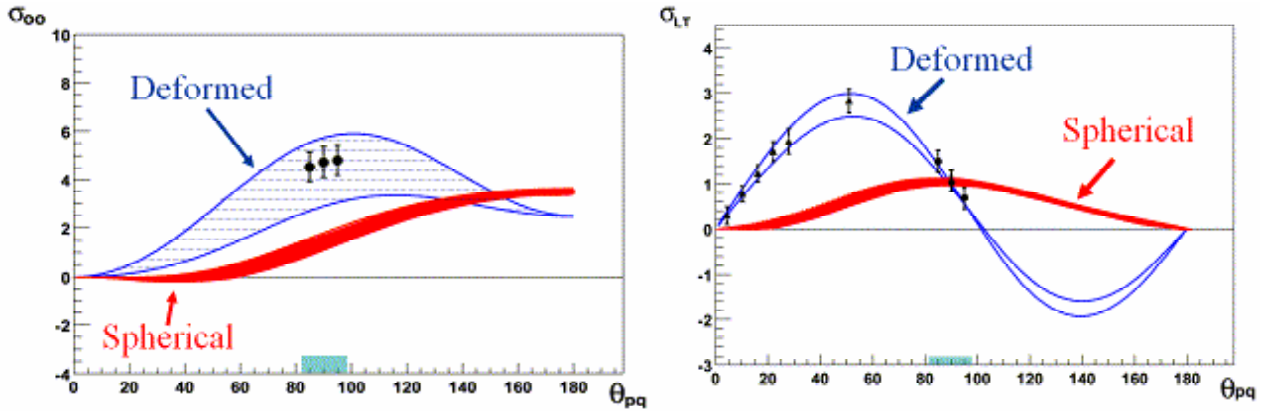}
  \caption{The currently employed methods of multipole extraction allow a
  reasonable determination of statistical and systematic
  uncertainties and an imprecise determination of model error,
  which often dominates.
  The error bands shown on the partial cross sections
  $\sigma_{LT}$ and $\sigma_{oo} \equiv \sigma_{E2}$
  depict the estimated model error with or
  without the assumption of spherical nucleon for the
  $N\rightarrow \Delta$ transition. While the conjecture of "deformation"
  is confirmed, understanding its origins will require a precise
  understanding and reduction of the model uncertainties
  (adopted from Ref~\cite{cnp03}).}
  \label{fig:rltband}
\end{figure}

The currently employed methodology for extracting multipoles has
served the field well, yielding important results.
Figure~\ref{fig:rltband} taken from reference~\cite{cnp03}, in which
Bates data~\cite{mertz01, kun03} are compared with theories that
account for them, demonstrates that the assumption of sphericity for
both the nucleon and the $\Delta^{+}(1232)$ is incompatible with the
data. The "deformed" band  is mapped by the spread of the model
predictions that successfully account for the data, while the dark,
"spherical", band spans the predictions of the same models in which
"sphericity" is imposed~\cite{cnp03}. However, investigating the
physical origin of deformation will require the comparison of
theoretical results which lie within the uncertainty band to the
data. A precise, quantitative definition of uncertainties of both
the experimentally derived and the theoretically produced values for
quantities such as $R_{LT}$ and of the multipoles that contribute to
it, is obviously required. The work presented partly addresses this
issue by presenting a method for extracting multipole information
from experimental nucleon resonance data in a rigorous, precise, and
model independent way.


\section{RESONANCE MULTIPOLE EXTRACTION}

It has been assumed up to now that to extract multipole
information  model independently from nucleon resonance data
(cross sections and polarization asymmetries), a complete set of
experimental observables is required.  Current experiments which
measure an incomplete set of observables typically rely on model
extraction. Multipoles have been presented in the literature
extracted by employing one of the following two approaches: a)
The Truncated Multipole Expansion (TME) approximation where most
or all of the non resonant multipoles are neglected (e.g.
see~\cite{frol99,kalleicher}) assuming that at the peak of the
resonance only resonant amplitudes contribute significantly and b)
The Model Dependent Extraction (MDE) method where certain
multipole amplitudes, often the resonant amplitudes, within a
phenomenological model description are adjusted to best describe
the data (e.g. see~\cite{mertz01,frol99,sp05}).  The second
method, MDE, is obviously superior to TME, for it assumes that
multipoles that cannot be determined from the data are fixed
through a model and not simply ignored. However it suffers from
the fact that the extracted values are biased by the model and
therefore characterized by a hard to evaluate systematic model
uncertainty. To ameliorate this deficiency, an Ansatz has been
proposed~\cite{cnp03} which is often used~\cite{sp05,stave06}
whereby the same data are analyzed employing different models
which describe the data adequately, and attributing the resulting
spread in the extracted quantities to model uncertainty. A
critical assessment of the this approach in the case of the
$N\rightarrow \Delta$ transition and the resulting uncertainties
is presented in reference~\cite{SBN_SOH}.

We have argued that the MDE method of extraction has up to now
served the field well; however, the refined questions that are
now emerging, such as the detailed understanding of the origin of
deformation,  require the development of a methodology which
addresses the following limitations of MDE:
\begin{itemize}
  \item A preconceived and somewhat arbitrary choice of the multipoles
  to extracted is required.
  \item The extracted values are model dependent and the model error
   is hard to quantify.
  \item Extraction of mutlipole values requires the existence of a
  model that successful describes the given resonance and the estimation
  of model error requires a multitude of them.
  \item Multipole amplitudes to which the experimental data are only
  moderately sensitive are impossible to extract.
\end{itemize}
A Model Independent Analysis Scheme, "AMIAS", is described here
which addresses most of the above mentioned limitations. Although
it is presented in this paper for the case of the $N\rightarrow
\Delta$ transition, it provides a framework and a methodology of
analysis which is applicable to all nucleon resonances. In the
sections that follow we shall develop the methodology and we will
subsequently apply it to the re-analysis of the $Q^{2}=0.127 \
GeV^2/c^2$  Bates and Mainz $N\rightarrow \Delta$  data. We will
conclude by discussing the advantages of the new method and the
possibilities that it offers which they need to be explored and
researched.

\subsection{The AMIAS Method and Postulate}
The AMIAS method has general applicability in a variety of
physical problems including the analysis of electroproduction data
in the nucleon resonance region. The method has a minimal
requirement that the parameters to be determined are linked in an
explicit way to the measured experimental quantities. There is no
requirement that this set of parameters provide an orthogonal
basis. Moreover these parameters can be subjected to explicit
constraints.

The essence of the method derives from the the following
postulate:

\begin{description}
  \item[Postulate:] A set of parameters $A_1^{j},A_2^{j},...,A_N^{j} =\{A_{\nu}\}^{j}$
  which completely and
  explicitly describes a physical process can be determined from
  relevant experimental observables $O^{M}=\{\textit{V}^{M}_{k} \pm \varepsilon^{M}_{k}\}$,
  (value  $\pm$ uncertainty), by assuming that
  any set $\{A_{\nu}\}^{j}$ constitutes a solution having a probability
  $P(j,M)$ of representing the "correct" solution which is equal to
    $$P(j,M)= erf[\chi^{2}(j,M)(A^{j}_{\nu},O^{M}_{k})]$$
    where
     $$\chi^{2}(j,M)=
     \sum_{k}\{\frac{(U^{j}_{k}-V^{M}_{k})}{\varepsilon^{M}_{k}}\}^{2}$$
  Thus $P(j,M)$ is a function of the $\chi^{2}$ resulting by the
  comparison to the experimental data $O^{M}=\{V_{k}, \varepsilon_{k}\}$
  of the predicted values $U^{j}_{k}$ by the $\{A_{\nu}\}^{j}$ solution.
\end{description}

We call an ensemble $Z^{M}$ of such solutions \emph{Canonical
Ensemble of Solutions}, which has properties that depend only on
the experimental data set $O^{M}$. Similarly a
\emph{Microcanonical Ensemble of Solutions} can be defined as the
collection of solutions which are characterized by
$$\chi^{2}\leq \chi^{2}_{min}+C $$ where C is usually taken to
be the constant equal to the effective degrees of freedom of the problem.

In the sections that follow we further develop the methodology
through the specific problem  of extracting multipole values from
nucleon resonances electroproduction data.

\subsection{Multipole Extraction employing AMIAS}
 In applying AMIAS to the problem of multipole extraction in the
$N\rightarrow \Delta$ transition the parameters to be extracted,
$\{A_{\nu}\}^{j}$ in the general formulation, are the multipole
amplitudes $M_{L\pm}$,  and if the data allow, the isospin
separated amplitudes $M^{1/2}_{L\pm}, M^{3/2}_{L\pm}$. The
experimental observables (data) \{$O_{i}$\} are typically cross
section and polarization asymmetries. Furthermore, the parameters
of our problem, the multipole amplitudes, are subjected to the
constraint of unitarization, by imposing the Fermi Watson theorem.
\begin{figure}[h]
  \includegraphics[height=8cm]{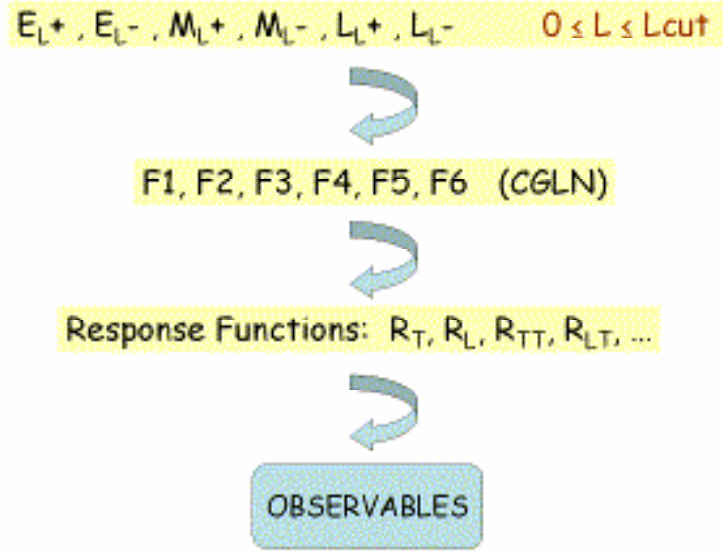}
  \caption{Multipole amplitudes, although not providing an orthogonal
   basis for describing the observables of electroproduction experiments,
   cross sections and polarization asymmetries, they provide a
   direct, complete and explicit parametrization linked via the
   CGLN amplitudes to the observables.}
  \label{fig:cgln}
\end{figure}
The linkage of observables to multipoles in the case of nucleon
resonances is shown schematically in Figure~\ref{fig:cgln}.  In
the  demonstration case we will examine which concerns the Bates-Mainz
data, the multipoles used refer to the $\pi^0p$ charge
channel which are connected to the $(A_p^{1/2},\ A_n^{1/2},\
A^{3/2})$ isospin-set through the relation:
$$A_{\pi^0p} = A_p^{1/2} + \frac{2}{3} A^{3/2} $$
as described in \cite{obs}. Following the standard practice of
the field, we shall also consider the phases of the multipoles as
known (from $\pi N$ scattering, see
\cite{obs,maid00,noz90,said02}) with extreme precision and
therefore fixed. The above choices allow a meaningful comparison
with previous analyses and model results; however they are not
inherent to the scheme. It is quite easy to consider the phases,
as experimental parameters characterized by uncertainties or even
to perform a combined analysis of electroproduction and $\pi N$
scattering data.

The AMIAS method produces numerical results, i.e. determines the
multipole amplitudes and the associated uncertainties, from the
experimental data by examining the properties of the ensemble of
solutions, canonical or microcanonical. In implementing the above
stated postulate to construct ensembles of solutions, statistical
concepts and extensive use of computational techniques are
employed. In the sections that follow we employ data and
pseudodata concerning the $N\rightarrow \Delta$ transition. This
approach would have been impossible to employ few years earlier;
it is the nowadays readily available formidable computational
power that has made such a scheme possible.

\subsection{The Algorithm and the Ensemble of Solutions}
The AMIAS postulate
is implemented using a Monte Carlo technique. Given a specific
set "M" of experimental data \{$O^{M}_{k}$\} the following
procedure is taken, employing a formulation that connects
multipoles to experimental observables~\cite{obs} (see
Figure~\ref{fig:cgln}):
\begin{enumerate}
  \item Choose a maximum cutoff, $L_{cut}$, for the multipoles that
  are to be explored in the analysis. This choice implies that a
  set \{$A_{\nu}$\} of $\nu$  parameters constitutes a solution to our
  problem.
  \item Assume a particular range of "acceptable" values for each of the
  multipoles. These ranges define the "phase volume" which the
  Monte Carlo method explores.
  \item Within the domain of acceptable values,
  randomly select a value (with equal probability) for the
  particular multipole. The randomly chosen values constitute
  a solution, the $j^{th}$ solution, $\{A_{\nu}\}^{j}$, to the
  problem.
  \item Using the data calculate the particular
  value of the $\chi^{2}(A_{\nu},O_{k})^{j}$ corresponding to the
  particular solution $\{A_{\nu}\}^{j}$.
  \item Repeating the above steps (1-4) a large number, typically a few
  hundred thousand, of solutions are generated which taken
  collectively they constitute the ensemble of solutions to our
  problem.
\end{enumerate}

\begin{figure}[h]
  \includegraphics[height=7.8cm]{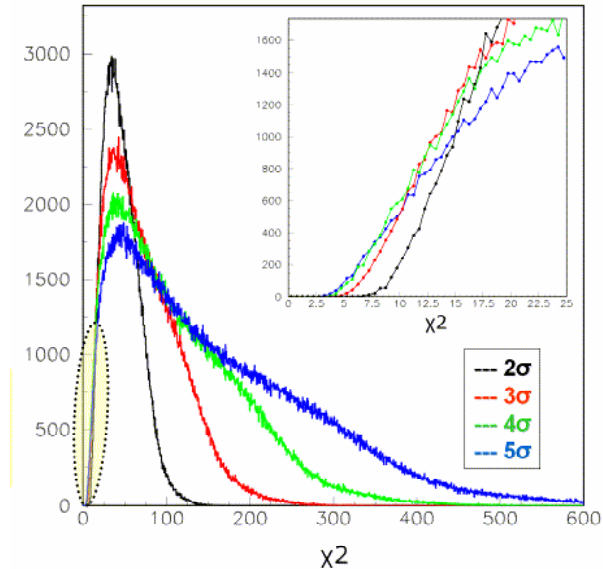}
  \caption{Typical histogram of solutions that is generated by randomly
   varying the model parameters $\{A_{\nu}\}$ (multipoles) within a given phase volume. It can be
   seen that, as the phase volume  increases, more "bad" solutions are generated
   while the distribution of "good solutions" (area in the dotted ellipse and
   insert), gets saturated. "$\sigma$" represents a unit of length in the phase
   volume and "good" or "bad" solutions refer to the $\chi^{2}$ value of the
   solution (see discussion in text)}
   \label{fig:chisqdistr}
\end{figure}

The method is totally insensitive to the choice of the multipole
cutoff $\ L_{cut}$ which should be chosen "sufficiently high" so
that the derived results do not change if $\ L_{cut}+1$ is used as
a maximum cutoff instead. The computational time required
increases roughly quadratically to the number of multipoles being
explored.

 The results presented below are derived by using
$L_{cut}=5$. Similarly the size of the phase volume that the
Monte Carlo method is sampling does not affect the solution
provided that the volume is "sufficiently large" to include all
"good" solutions. By "good solutions" we denote solutions with
small or reasonable $\chi^{2}$ values as compared to the degrees
of freedom of the problem. The determination of the phase volume
might require successive attempts so it is appropriately bounded
or the employment of self adjusting algorithms, as will comment
below. The computational time required depends critically on a
"reasonable" choice for the phase volume to be explored.

Each one of the solutions in the ensemble $\{A_{\nu}\}^{j}$ is
tagged by a $(\chi^{2})^{j}$  value. It is obvious that the
overwhelming majority of the solutions generated in this fashion
are characterized by very large $\chi^{2}$  values. This can be
seen in the histogram of  $\chi^{2}$  values which result for a
given choice of the phase volume shown in
Figure~\ref{fig:chisqdistr}. The behavior of the distribution is
altered by changing the "size" of the phase volume. To visualize
the effect, we have defined a scale "$\sigma$" within which a
given multipole is allowed to vary. We subsequently have
increased the phase volume explored by increasing the range
within which multipole is allowed to vary by multiples of this
basic scale. As the volume is increased, the distribution is
changed, basically by producing more solutions characterized by
terrible $\chi^{2}$  values, as expected. However, as is also
evident in the insert, a saturation of the "good values" is
achieved; we simply are not able to identify new regions of good
solutions by exploring wider range of values for the multipoles.

\begin{figure}[h]
  \includegraphics[height=10cm]{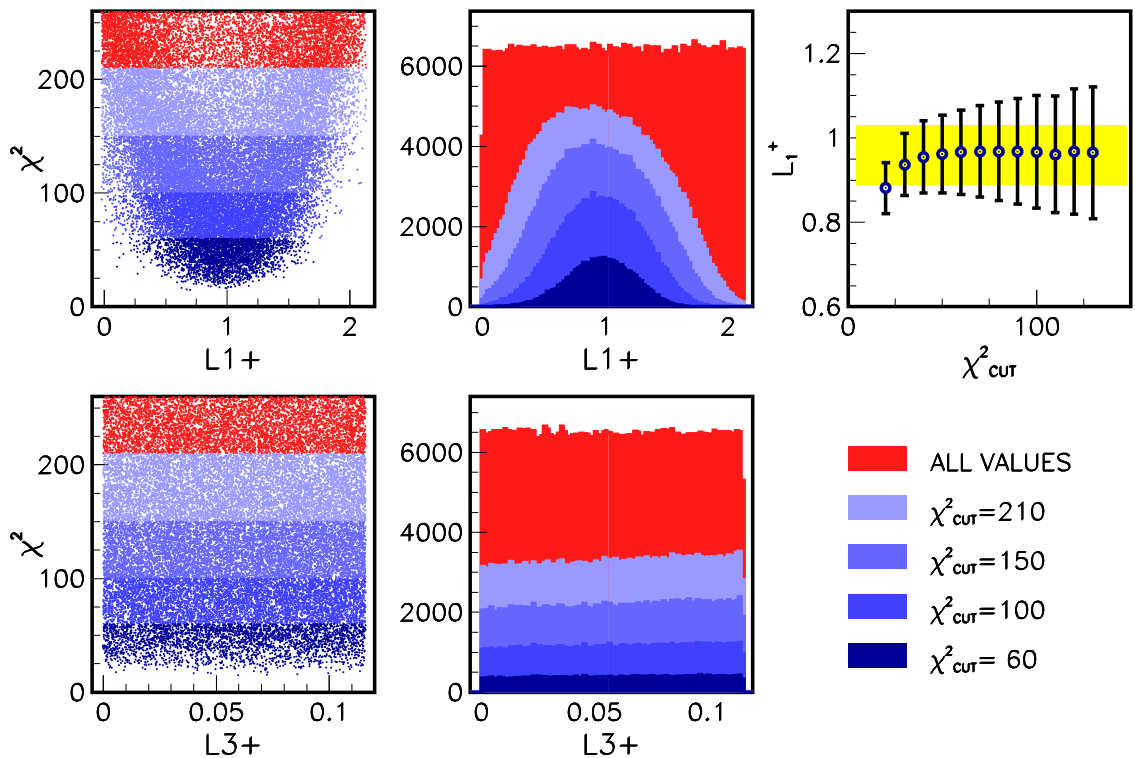}
  \caption{The ensemble of solutions that is generated by randomly
   varying the parameters $A_{\nu}$ (multipoles) can be visualized by projecting
   it on a two-dimensional surface ordered according to its $\chi^{2}$ value for
   a given multipole, shown in the left panels. In the middle panels,
   histograms of the populations that results by applying
   successive cuts on the $\chi^{2}$ values are depicted. On the righthand
   panel the dependency of the derived central values and widths of the resulting
   distributions are shown.}
   \label{fig:chisqensemble_a}
\end{figure}

The behavior of a given model parameter $A_{\nu}$ (multipole) in
the ensemble of solutions can be visualized by projecting the
solutions on a two dimensional surface where the value of the
parameter is projected in the ordinate and its corresponding
$\chi^{2}$ value on the abscissa. This is done in the left column
of Figure~\ref{fig:chisqensemble_a}. The top row concerns the
$L_{1+}$ multipole, for which the data are known to be
particularly sensitive, while the bottom row that of the the
$L_{3+}$ multipole to which the particular data set exhibits
little or no sensitivity.
For a specific value of the multipole the ensemble contains
solutions characterized by a wide range of $\chi^{2}$ values
depending of the choice of the other multipoles whose values are
not visualized. For a given value of the multipole it can be seen
that a solution can be identified which is characterized by a
minimum value of $\chi^{2}$. For the $L_{1+}$ multipole, but not
for the $L_{3+}$, the observed minima show strong dependence on
the value of the multipole, thus manifesting its sensitivity to
the data. This sensitivity becomes explicit and quantifiable by
applying successive cuts on the $\chi^{2}$ values and
constructing out of the selected population of solutions
histograms, shown in the middle column of the figure. It can be
seen that the distribution of solutions is uniform (flat) if the
entire population is examined, manifesting the randomness of the
choice (no model bias). However, as we progressively limit the
solutions accepted by restricting the acceptable $\chi^{2}$
values, a distribution of values with a well defined maximum and
width emerges, but only for the ($L_{1+}$) multipole, to which
the data are sensitive; for the non sensitive ($L_{3+}$), no
distribution with a finite width can be found. Finally, in the
righthand panel of the top row of
Figure~\ref{fig:chisqensemble_a} the dependence on the values of
the maxima and of the widths of the resulting peaks is explored
as a function of $\chi^{2}_{cut}$ . A stable saturation value for
the peak position emerges but not for its width which grows as a
function of $\chi^{2}_{cut}$.

Certain features emerge, which are found to hold for all
multipoles, not only $L_{1+}$ and $L_{3+}$: The peak of the
distribution is highly insensitive to the $\chi^{2}$ cut while the
width grows as a function of the $\chi^{2}$ cut. The method does
not treat differently "sensitive" from "non sensitive"
multipoles; they naturally emerge as such. The widths of the
histograms, ranging from very narrow to very wide or infinite
widths, naturally select and order the various multipoles
according to their sensitivity to the data set.

\subsection{Correlations}
A central issue  of the problem of extracting multipole amplitudes
which is properly treated in the AMIAS method, is the handling of
correlations. Both the TME and the MDE methods, by freezing all
the "insensitive" multipoles,  exclude any possibility of
determining them. In addition the derived solution has in it
embedded the correlations to those multipoles; the result is that
this source of model error is reflected both by "shifting" the
extracted values for the dominant amplitudes and by
underestimating the model uncertainty.

\begin{figure}[h]
  \includegraphics[height=8cm]{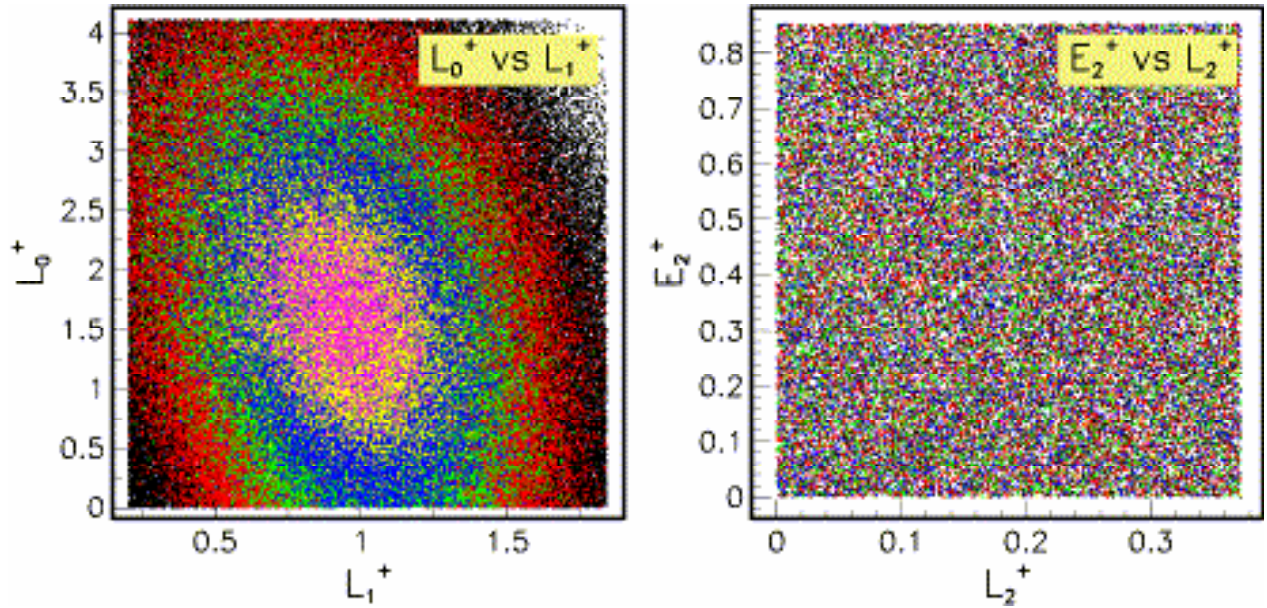}
  \caption{The two-dimensional scatter plot in which the ensemble
  of solutions is projected on the plane defined by their values
  and color coded according to the $\chi^{2}$ value,
  provides an easy way to visualize correlations.
  In the left panel of the figure the strong correlation between the
  $L_{0+}$ and $L_{1+}$ multipoles is obvious.
  The right panel does not indicate any correlation between
  the $E_{2+}$ and $L_{2+}$ multipoles.}
  \label{fig:correlations}
\end{figure}

The AMIAS method, addresses  these shortcomings and it provides
for an easy visualization of the existence of such correlations.
All possible correlations are accounted by allowing all multipoles
to randomly vary, and to yield solutions with all allowed values
of the "insensitive" multipoles. The visualization of at least the
dominant correlations is accomplished in a two-dimensional
scatter plot in which the ensemble of solutions is projected on
the plane defined by their values and color coded according to
the $\chi^{2}$ value. In the left panel of
Figure~\ref{fig:correlations} the strong correlation between the
$L_{0+}$ and $L_{1+}$ multipoles is obvious. It is worth pointing
out that in the most recent analysis of $N\rightarrow \Delta$
data the $L_{0+}$ is frozen  at the model
values~\cite{sp05,stave06}.  The right panel explores the
correlation between the $E_{2+}$ and $L_{2+}$ multipoles which
visually is shown to be absent.

\subsection{Probability Distributions}
\begin{figure}[h]
  \includegraphics[height=10cm]{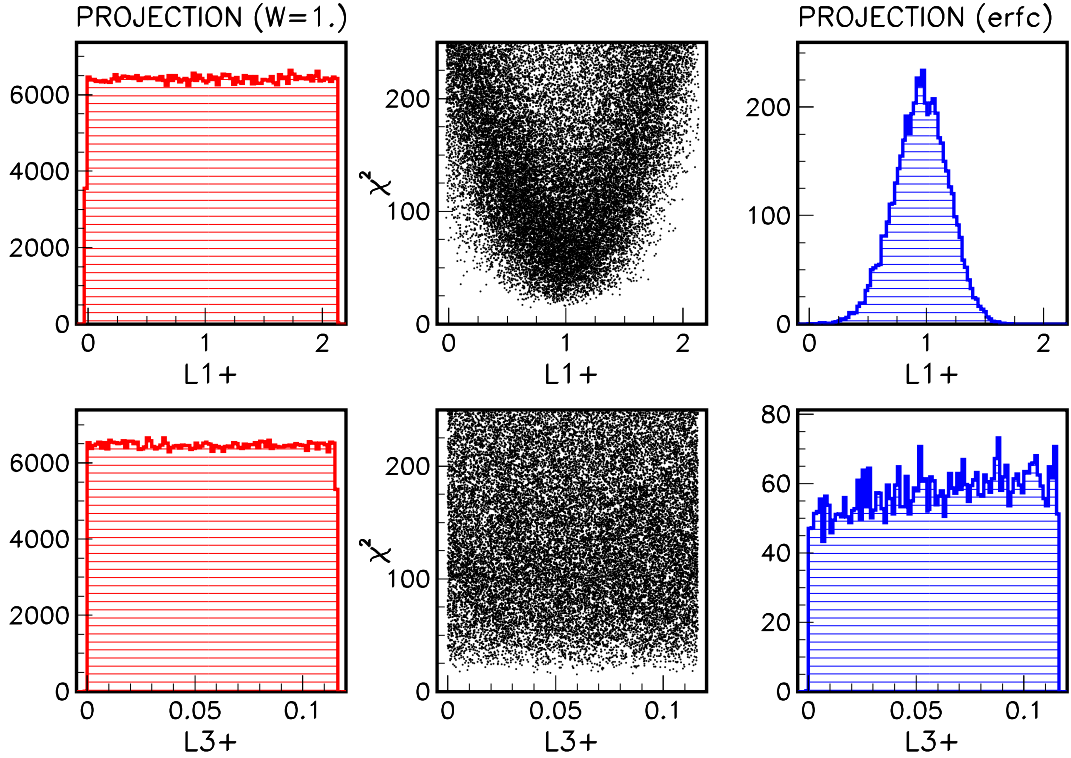}
  \caption{The two-dimensional projections of the ensemble of solution on a
  $A_{\nu}~ \chi^{2}$ plane, same as those of
  Figure~\ref{fig:chisqensemble_a},
  are shown in the middle panels. They are mapped into probability
  distributions by implementing the AMIAS postulate
  (right hand panels), or  to flat distributions (left hand panels) if an equal probability
  is assumed manifesting the model independence of the method.}
  \label{fig:weighteddistibution}
\end{figure}

The results that emerge by examining the behavior of the ensemble
of solutions provide an understanding of the method, but they do
not make use of the AMIAS postulate. Implementing the postulate
allows the extraction of values \emph{and} uncertainties from the
data. The following procedure is followed: For each solution of
the ensemble $\{A_{\nu}\}^{j}$ we take the corresponding value of
the multipole of interest $A_{\nu}^{j}$ and we assign to it a
probability, as our postulate requires, equal to $P(j,k)=
erf[\chi^{2}(A^{j}_{\nu},O_{k})]$. We then form a histogram of
the multipole values of the ensemble weighted by the probability
$P(j,k)$. The resulting distributions, shown in the right hand
panel of the Figure~\ref{fig:weighteddistibution}, are
characterized by a peak position and a width; the histogram for
$A_{\nu}=L_{1+}$ has a well defined maximum and a well defined
width while the corresponding histogram for $A_{\nu}=L_{3+}$ is
featureless manifesting its insensitivity to the data. The
resulting distribution is the relative probability distribution
for this particular parameter (multipole) to represent the
physical value. We take its mean to be the derived (extracted)
from the data value; this value is characterized by an
uncertainty which is completely determined by the probability
distribution. A numerical value for the uncertainty  can be
derived from the distribution for a given confidence level
(C.L.). For instance, a "$1 \sigma$" (68\% C.L.) value is given
by the $\pm A_{\nu}$ values which bound $\pm34\%$ of the integral.

In the subsequent section we will demonstrate through the use of
pseudodata, for which the generating multipoles are precisely
known, that the method can reproduce them with a well understood
uncertainty.

\subsection{Validation with Pseudodata}

AMIAS produces  values and uncertainties for multipoles for which
the data exhibit sensitivity, in a manifestly model independent
manner. The values and uncertainties, have been demonstrated to
have a precise meaning through the analysis of pseudodata which
were generated with predetermined statistical accuracy. We have
used as generator the MAID code, with our particular set of input
multipole parameters. In the cases presented below, we have
frozen the $A_{\nu}^{1/2}$  and we have varied the
$A_{\nu}^{3/2}$ amplitudes. Few demonstrative cases of the
pseudodata validation are presented below.

The demonstration case presented here is based on the kinematics
of the $Q^{2}=0.127$ (GeV/c)$^2$~  Bates and Mainz $N\rightarrow
\Delta$ data. The data set is published, is well understood and
it is well described by MAID if the resonant amplitudes are
adjusted~\cite{sp05}. Several sets of pseudodata were generated,
at identical kinematics to those of the experimental values, in
order to test the ability of the AMIAS code to reproduce  a) the
generator multipole values and b) the input uncertainties imposed
on the generator multipoles.

\begin{table}[h]
\caption{The multipole values extracted, in units of $10^{-3}/
m_{\pi}$, from the three pseudodata sets are compared to the
generator ( modified MAID) values. They are shown to be entirely
compatible with increasing accuracy as required.}
\label{tab:pseudovalues}
\begin{tabular}{|c|r|r|r|r|}
  \hline
  Multipole &     Generator  & Set A $\quad\quad  $ & Set B $\quad\quad  $ & Set C $\quad\quad  $ \\
  \hline
  $M_{1+}$  & $27.248\ $ & $27.23\ \pm\ 0.13\ $ & $27.229\ \pm\ 0.013$ & $27.249\ \pm\ 0.001$ \\
  $L_{0+}$  & $ 3.500\ $ & $ 3.70\ \pm\ 0.23\ $ & $ 3.515\ \pm\ 0.022$ & $ 3.502\ \pm\ 0.002$ \\
  $L_{1+}$  & $ 1.048\ $ & $ 1.03\ \pm\ 0.08\ $ & $ 1.047\ \pm\ 0.008$ & $ 1.048\ \pm\ 0.001$ \\
  $E_{1+}$  & $ 1.481\ $ & $ 1.49\ \pm\ 0.18\ $ & $ 1.489\ \pm\ 0.017$ & $ 1.482\ \pm\ 0.002$ \\
  $E_{0+}$  & $ 4.225\ $ & $ 3.68\ \pm\ 1.02\ $ & $ 4.278\ \pm\ 0.135$ & $ 4.239\ \pm\ 0.013$ \\
  $M_{1-}$  & $ 4.119\ $ & $ 4.47\ \pm\ 1.31\ $ & $ 4.161\ \pm\ 0.126$ & $ 4.124\ \pm\ 0.013$ \\
  $L_{1-}$  & $ 1.205\ $ & $ 1.05\ \pm\ 0.43\ $ & $ 1.170\ \pm\ 0.080$ & $ 1.203\ \pm\ 0.008$ \\
  $E_{2-}$  & $ 1.024\ $ & $ 1.07\ \pm\ 0.45\ $ & $ 1.053\ \pm\ 0.061$ & $ 1.027\ \pm\ 0.006$ \\
  $L_{2+}$  & $ 0.007\ $ & $ 0.02\ \pm\ 0.01\ $ & $ 0.008\ \pm\ 0.001$ & $ 0.008\ \pm\ 0.001$ \\
  $E_{2+}$  & $ 0.006\ $ & $ 0.01\ \pm\ 0.01\ $ & $ 0.009\ \pm\ 0.001$ & $ 0.007\ \pm\ 0.001$ \\
  \hline
\end{tabular}
\end{table}

The validation of the ability of the method to reproduce the
central values is straight forward. Three sets of pseudodata were
generated, characterized by the increasing statistical accuracy:
i) "Set A" with statistical accuracy similar to that of the
experimental values ii) "Set B" with statistical accuracy ten
times better than that of the experimental values and iii) "Set
C" with statistical accuracy hundred times better than that of
the experimental values. These data were analyzed and the
multipoles were extracted which are tabulated and compared with
the generator values in Table~\ref{tab:pseudovalues}. We have
tabulated only extracted values which are derived with
uncertainties better than 100\% for "Set A". It can be seen that
the multipole values are extracted accurately, in complete
agreement with the generator values within the stated statistical
accuracy. Also, as required, the quoted uncertainties are reduced
(tenfold and hundredfold), proportionally to the statistical
accuracy of the pseudodata sets.

\begin{figure}[h]
  \includegraphics[height=5.8cm]{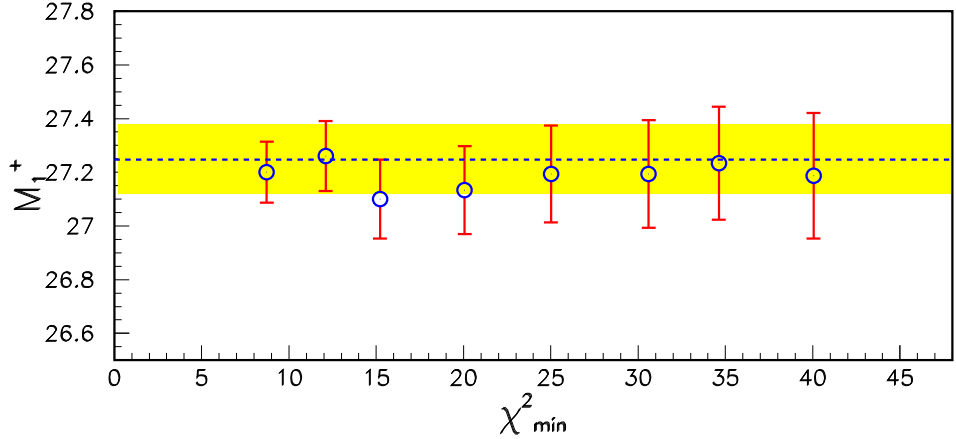}
  \caption{The derived values and uncertainties for the $M_{1+}$
  multipole as a function of the $\chi^{2}$ value of the pseudodata
  that were generated. The shaded area around the nominal value depicts the
  $1\sigma$ variation which was imposed on this multipole as a "model uncertainty".}
  \label{fig:pseudoerrors}
\end{figure}

To test the ability of the method to extract uncertainties which
have precise statistical interpretation is generally more
difficult. The scaling behavior exhibited by the three sets of
pseudodata, discussed above, is a necessary but not sufficient
condition.  The definitive validation was achieved by introducing
an arbitrary uncertainty, a "model uncertainty" to the nominal
generator values.

The introduction of a model uncertainty allows the generation of
pseudodata  for which the uncertainties are defined by the model
uncertainty and not the statistical precision of the data set
(corresponding to the limit of infinite statistics). Pseudodata
with increasing $\chi^{2}$ values, depending on the magnitude of
the model error, were analyzed using the AMIAS method. AMIAS is
demonstrated to reproduce the central value of a given multipole
of the generator and to assign to it an uncertainty which is
statistically compatible to the predetermined "model error". This
is shown for the case of the $M_{1+}$ multipole in
Figure~\ref{fig:pseudoerrors}.

The detailed investigation of the method using pseudodata, which
is far more extensive than is documented in this article,
demonstrated that AMIAS, unlike TME and MDE provides for the
first time a methodology and a tool for identifying and  precisely
extracting multipole amplitudes in a model independent way. In
addition the quoted uncertainties  have a precise, well
understood meaning. We proceed in the following section to apply
the method to reanalyze the previously published $Q^2=0.127
GeV^2/c^2$ data.

\section{ Example: Re-analysis of the $Q^2=0.127$  $GeV^2/c^2$  data}

To demonstrate the capabilities of the method we re-analyzed the
the $H(e,e'p)\pi^0$ measurements performed at $Q^2=0.127$
$GeV^2/c^2$ and $W=1232$ MeV. This set consists of Bates and
Mainz data; a detailed description and analysis of the data can
be found in \cite{sp_SOH}. The data set consists of 31 data
points, cross section results for the $\sigma_{TT}$,
$\sigma_{LT}$, $\sigma_0$, $\sigma_{E2}$ and the polarized beam
cross section $\sigma_{LT'}$. Since the MAID model \cite{maid00}
provides for this $Q^2$ a good description of the data set, the
MAID-2003 multipoles are used as a starting point for the AMIAS
method. As commented earlier, the starting point can be
arbitrary, however a good starting point provides easy
convergence and considerable savings in computer time. A
$L_{cut}=5$ value was chosen, so that a sufficiently large number
of background amplitudes are included in the computational
exploration. Results are derived by uniformly varying the
real and imaginary part of the input amplitudes of the model
in the $\pi^0$ charge-channel after a reiterative
selection of the phase volume to be explored. The unitary box
width $w_0$ assigned to each of the input amplitudes was set to
the $\pm 10\%$ of their central (MAID) value. Amplitudes in
general were allowed to vary normally in a range of $20\times w_0$;
this range was reduced for the sensitive  $E_{1+}$ and
$L_{1+}$ amplitudes and particularly for the $M_{1+}$ multipole.

\begin{figure}[h]
  \includegraphics[height=8cm]{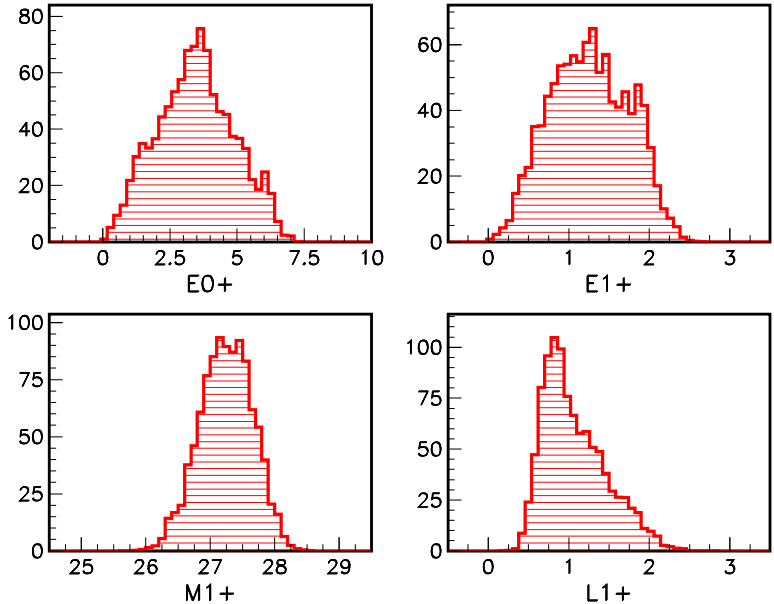}
  \caption{Probability distributions for the norms of some of the sensitive
  amplitudes of the analyzed Bates/Mainz data set. The distributions allow
  the determination of the central value and corresponding uncertainty for
  each of the multipoles.}
  \label{fig:bates_sens_ampl}
\end{figure}

The resulting (un-normalized) probability distributions  for some
of the sensitive amplitudes are shown in
Figure~\ref{fig:bates_sens_ampl}. The probability distributions,
as in the case of the pseudodata, allow the determination of both
the value and the uncertainty of each of the sensitive multipoles.
The extracted multipole values and ($1\sigma$ confidence)
uncertainties, of the Bates/Mainz data fit with the AMIAS method
are presented in the Table~\ref{tab:bates_results}.

\begin{table}[h]
\caption{Extracted values and ($1\sigma$) uncertainties for the
norm of the sensitive multipoles  from the Bates/Mainz
experimental data set using the AMIAS method. The results are
presented with decreasing multipole sensitivity, which is
reflected in the relative uncertainty. The MAID-2003, SL and DMT
model values are tabulated for comparison. The multipoles are in
units of $10^{-3}/ m_{\pi}$ }. \label{tab:bates_results}

\begin{tabular}{|c|r|r|r|r|r|}
\hline
  Multipole & Extracted Value           & Relative Error    & MAID-2003       & Sato \& Lee     & DMT $\quad$    \\
\hline
  $M_{1+}$  & $ 27.24\ \pm\ 0.20       \quad$ & $  0.73\ \%\quad$ & $ 27.464 \quad$ & $ 27.661 \quad$ & $ 27.489 \quad$\\
  $L_{1+}$  & $  0.82\ ^{+0.20}_{-0.09}\quad$ & $ 17.7 \ \%\quad$ & $  1.000 \quad$ & $  0.672 \quad$ & $  0.986 \quad$\\
  $L_{0+}$  & $  2.23\ \pm\ 0.41       \quad$ & $ 18.4 \ \%\quad$ & $  2.345 \quad$ & $  1.008 \quad$ & $  1.994 \quad$\\
  $E_{0+}$  & $  3.44\ \pm\ 0.70       \quad$ & $ 20.3 \ \%\quad$ & $  2.873 \quad$ & $  2.213 \quad$ & $  3.206 \quad$\\
  $E_{1+}$  & $  1.16\ ^{+0.32}_{-0.24}\quad$ & $ 24.1 \ \%\quad$ & $  1.294 \quad$ & $  1.288 \quad$ & $  1.401 \quad$\\
\hline
\end{tabular}
\end{table}

The quoted uncertainties for the extracted multipoles include all
experimental errors (statistical + systematic) and obviously
contain no model error. The relative uncertainties, which can be
considered as a measure of the sensitivity for each of the
extracted multipoles, are also tabulated. The values of the
MAID-2003, Sato \& Lee \cite{sato} and DMT \cite{kama,dmt00}
model are also given in the same table for comparison. It is
important to highlight the fact that the AMIAS method yields
information on five multipoles, while the MDE method only for the
three resonant amplitudes~\cite{SBN_SOH, sp_SOH}.

\begin{figure}[h]
  \includegraphics[height=11cm]{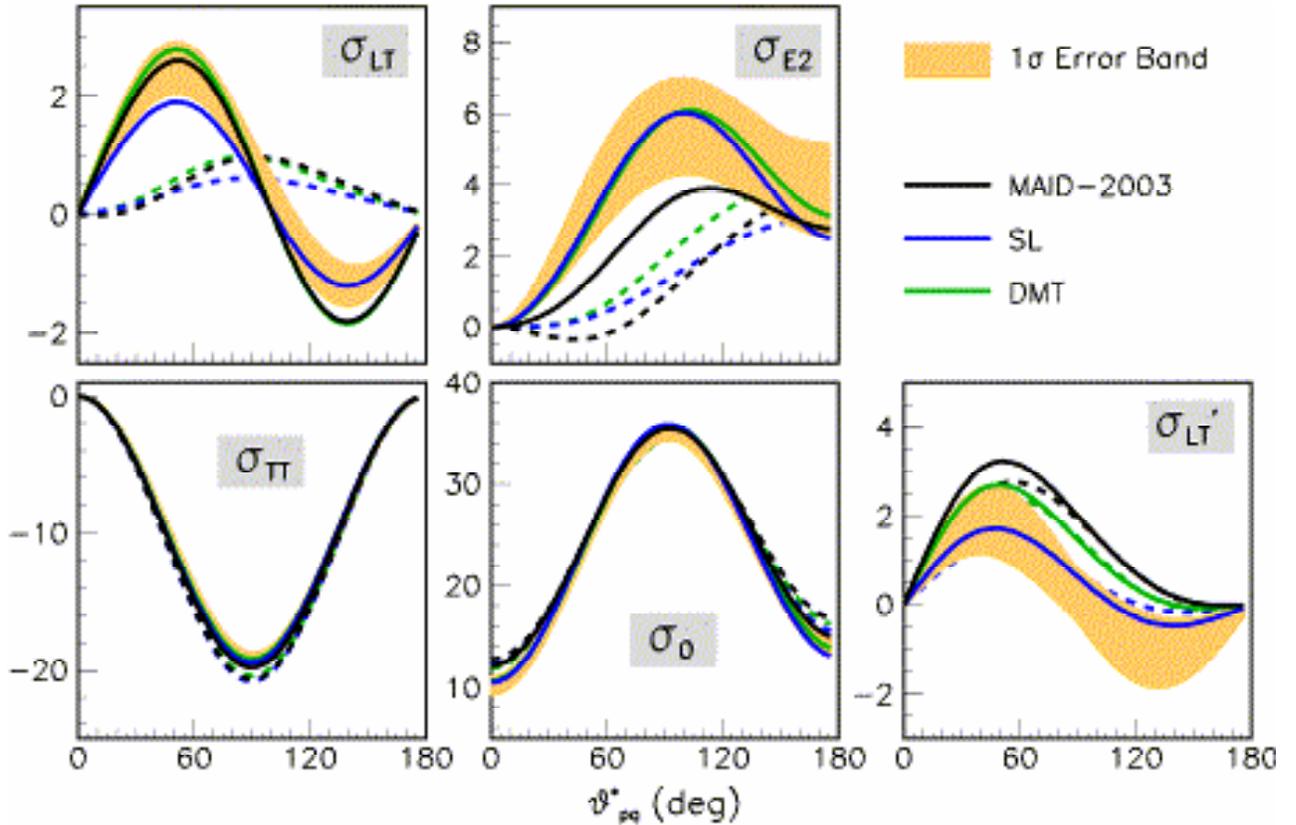}
  \caption{The experimentally allowed region for the partial cross sections
  for $Q^2=0.127$ $GeV^2/c^2$ and $W=1232$ MeV, as determined by the analysis
  of the Bates//Mainz data by AMIAS: the orange (shaded) bands
  depict the allowed region with $1\sigma$ confidence.
  The solid curves show the model predictions while the dashed curves show the "spherical"
  solutions of the same models.}
  \label{fig:bates_bands}
\end{figure}

In Figure~\ref{fig:bates_bands} the allowed region for partial
cross sections for $Q^2=0.127$ $GeV^2/c^2$ and $W=1232$ MeV is
shown in the shaded area (bands). These bands are defined through
the result of the AMIAS solution for  the Bates/Mainz data set.
The shaded band shows the envelope that accommodates all possible
solutions that are compatible with the experimental data with
$1\sigma$ (68\%) confidence level. These uncertainty bands are
model independent.

The error bands of Figure~\ref{fig:bates_bands} are solely
defined by the experimental data, with the minimal assumptions of
the AMIAS method, spelled out earlier. In contrast, the bands in
Figure~\ref{fig:rltband} are  model defined, representing the
spread of the model solutions to the same data that "acceptable"
models offer at a given time. It can be observed that the
experimentally allowed region is broader and covers a slightly
different region that that covered by the model bands of
Figure~\ref{fig:rltband}.

The predictions of the  DMT, MAID and SL models are also shown
in Figure~\ref{fig:bates_bands} (solid curves) along with the
corresponding results for their "spherical" solutions (dashed
curves, same color coding). It can be seen that the model
solutions fall within the experimentally allowed region. As
expected, only the partial cross sections $\sigma_{LT}$ and
$\sigma_{E2}$ discriminate, but with a high level of confidence,
between the deformed and spherical solutions.

A model independent isolation of resonant quadrupole strength
$(M^{3/2}_{1+},E^{3/2}_{1+}, L^{3/2}_{1+} )$ and an extraction of
the EMR and CMR values is allowed by AMIAS, if the data can
support such a separation. Polarization and isospin sensitive
observables are required. Such data are emerging from the new CLAS
measurements, where both $(e,e' \pi^{+})$ and $(e,e'
\pi^{\circ})$ have been measured~\cite{Smith_SOH}.

\begin{table}[h]
\caption{Comparison of the extracted multipoles from the full and
a reduced data set, in which the $\sigma_{LT'}$ cross sections
have been excluded. The relative change in the derived values is
also indicated. The multipoles are in units of $10^{-3}/
m_{\pi}$}. \label{tab:bates_reduced}

\begin{tabular}{|c|r|r|r|}
\hline
            & Full Data Set   & Reduced Data Set &           \\
\hline
  Multipole & Extracted Value & Extracted Value  & Change    \\
\hline
  $M_{1+}$ & $ 27.24\ \pm\  0.20      \quad$ & $ 27.15 \ \pm\  0.24      \quad$ & $  +0.3\ \%\  $ \\
  $L_{1+}$ & $  0.82\ ^{+0.20}_{-0.09}\quad$ & $  0.90 \ ^{+0.30}_{-0.14}\quad$ & $  -9.8\ \%\  $ \\
  $L_{0+}$ & $  2.23\ \pm\ 0.41       \quad$ & $  2.63 \ \pm\  0.60      \quad$ & $ -18.0\ \%\  $ \\
  $E_{0+}$ & $  3.44\ \pm\ 0.70       \quad$ & $  3.74 \ \pm\  0.76      \quad$ & $  -8.7\ \%\  $ \\
  $E_{1+}$ & $  1.16\ ^{+0.32}_{-0.24}\quad$ & $  1.19 \ ^{+0.28}_{-0.20}\quad$ & $  -2.6\ \%\  $ \\
\hline
\end{tabular}
\end{table}

The ability of the AMIAS method to determine correctly and
selectively the multipole sensitivities  contained in a given data
set, was demonstrated by analyzing a subset of the Bates/Mainz
data set in which polarization observables, ($\sigma_{LT'})$, were
excluded. It is known that the $\sigma_{LT'}$ cross section
depends on the background amplitudes and particularly on the
$L_{0+}$. A new multipole solution was generated with AMIAS based
on the reduced data set. The derived results are listed in
Table~\ref{tab:bates_reduced} and compared to the values extracted
from the full data. The expected sensitivities are borne out as
can be seen in the last column in the table.  It is clear from
the derived values, that the change in the resonant amplitudes is
minimal and lies within the range of their extracted uncertainty.
As expected the sensitive background amplitudes, and especially
the $L_{0+}$, indicate a significant change in their initial
values.


\section{Summary, Conclusions and Future Work }
We have developed and presented aspects of a scheme for the
analysis of experimental data (AMIAS) which is very general and it
allows the extraction of maximal information from incomplete data
sets. The AMIAS method is shown to be suitable for the analysis of
nucleon resonance electroproduction data in general although it
was discussed only for the case of  the $N\rightarrow \Delta$
transition data. The new method offers significant advantages over
the currently employed methods and is demonstrated to:
\begin{itemize}
  \item Be model independent.
  \item Extract maximum information for all available multipoles,
  without any bias; it is capable of ranking them in order of
  significance.
  \item Account for the correlations amongst the contributing
  multipoles, and to provide an easy visualization of them.
  \item Yield uncertainties which have  a precise meaning, in terms of confidence levels.
  \item Be numerically robust, regardless of the data base.
\end{itemize}

The re analysis of the published and well understood $Q^{2}=0.127
GeV^2/c^2$ Bates and Mainz $N\rightarrow \Delta$ data was
presented as a demonstration case.  Information on five
multipoles is derived, instead of the three that were extracted
using the standard methodology. The derived values for $M_{1+}$,
$E_{1+}$ and $L_{1+}$  are compatible with the published
experimental values, but they are determined with higher
precision.

A number of developments are under way. A  W dependent analysis is
being currently developed. In parallel, the coding and the
algorithms are being improved in order to improve their
efficiency. A self adopting algorithm, which searches the
starting point and optimally bounds the phase space volume, is
also under investigation. A (re) analysis of the high quality
data sets at a number of $Q^{2}$ values that is now emerging is
planned.

The model independent, precise determination of the multipole
amplitudes and the quantification of the associated experimental
uncertainty provides a tool for a meaningful comparison of data
with competing interpretation schemes and models.  In the case of
the issue of hadron deformation such a precise and quantitative
comparison is need in order to understand the mechanisms that
generate it.


\begin{theacknowledgments}
We would like to acknowledge and thank  C.~Alexandrou,
A.~Bernstein, K. DeJager, T.S.~Lee, L. Tiator and M.~Vanderhaegen
for many interesting discussions on this subject and in
particular the first three for a critical review of the
manuscript. We would like also to acknowledge the technical
support of the IASA GRID-team for using the HG-02-IASA and
GR-06-IASA nodes. This work is partially supported by the program
PYTHAGORAS co-funded by the European Social Fund and National
Resources (EPEAEK II).

\end{theacknowledgments}


\end{document}